\documentclass{article}
\usepackage{spconf,amsmath,graphicx}
\usepackage{amssymb}
\usepackage{mathrsfs}
\usepackage[T1]{fontenc}

\title{Audio Set classification with attention model: A probabilistic perspective}
\name{Qiuqiang Kong*, Yong Xu*, Wenwu Wang, Mark D. Plumbley\thanks{* These first two authors contribute equally to this work.}}
\address{Center for Vision, Speech and Signal Processing, University of Surrey, UK}
\email{\{q.kong, yong.xu, w.wang, m.plumbley\}@surrey.ac.uk}

\begin{document}
\maketitle
\begin{abstract}
This paper investigates the classification of the Audio Set dataset. Audio Set is a large scale weakly labelled dataset of sound clips. Previous work used multiple instance learning (MIL) to classify weakly labelled data. In MIL, a bag consists of several instances, and a bag is labelled positive if at least one instances in the audio clip is positive. A bag is labelled negative if all the instances in the bag are negative. We propose an attention model to tackle the MIL problem and explain this attention model from a novel probabilistic perspective. We define a probability space on each bag, where each instance in the bag has a trainable probability measure for each class. Then the classification of a bag is the expectation of the classification output of the instances in the bag with respect to the learned probability measure. Experimental results show that our proposed attention model modeled by fully connected deep neural network obtains mAP of 0.327 on Audio Set dataset, outperforming the Google's baseline of 0.314 and recurrent neural network of 0.325. 
\end{abstract}
\section{Introduction}
Analysis of environmental sounds has been a popular topic which has the potential to be used in many applications, such as public security surveillance, smart homes, smart cars and health care monitoring. Audio classification has also attracted significant research effort due to the Detection and Classification of Acoustic Scenes and Events (DCASE) challenge \cite{mesaros2016tut, mesaros2017dcase}. Several tasks have been defined for audio classification including acoustic scene classification \cite{mesaros2016tut}, sound event detection \cite{mesaros2016tut} and audio tagging \cite{yongIJCNN2017, xu2017trans}. However, the data sets used in these challenges are relatively small. Recently, Google released an ontology and human-labeled large scale data set for audio events, namely, Audio Set \cite{gemmeke2017audio}. Audio Set consists of an expanding ontology of 527 sound event classes and a collection of over 2 million human-labeled 10-second sound clips drawn from YouTube videos.

Audio Set is defined for tasks such as audio tagging. The objective of audio tagging is to perform multi-label classification on fixed-length audio chunks (i.e. assigning zero or more labels to each audio chunk) without predicting the precise boundaries of acoustic events. This task was first proposed in DCASE2016 challenge. Deep neural networks (DNNs) \cite{gemmeke2017audio} and convolutional recurrent neural networks (CRNNs) \cite{yongIJCNN2017} have been used for predicting the occurring audio tags. Neural networks with an attention scheme was firstly proposed in our previous work \cite{yongIS2017} for the audio tagging task which provides the ability to localize the related audio events. Gated convolutional neural networks \cite{icassp2018_dcase2017} have also been applied in the ``Large-scale weakly supervised sound event detection for smart cars'' task of DCASE2017 challenge, where our system achieved the 1st place in the audio tagging sub-task\footnote{http://www.cs.tut.fi/sgn/arg/dcase2017/}. However, the audio tagging data set used in the DCASE2017 challenge is just a small sub-set of Google Audio Set \cite{gemmeke2017audio}. The number of the audio event classes is only 17 compared with 527 classes in Google Audio Set. In this paper, we propose to use an attention model for audio tagging on Google Audio Set \cite{gemmeke2017audio}, which shows better performance than the Google's baseline. In this work, we have two main contributions, one is that we conduct and explore a large-scale audio tagging on Google Audio Set \cite{gemmeke2017audio}. Secondly, we explain the attention model from a probability perspective. The attention scheme is also similar to the feature selection process which can figure out the related features while suppressing the unrelated background noise. It is achieved by a weighted sum over frames where the attention values are automatically learned by the neural network model.

In the remainder of this paper, the related works are presented in Section \ref{sec:related_work}. The proposed attention method and explanation from the probability perspective are shown in Section \ref{sec:attention}. Section \ref{sec:exp} presents the experimental setup and results. Conclusions are drawn in section \ref{sec:conc}.

\section{Related works} \label{sec:related_work}


Multiple instance learning (MIL) \cite{Maron1998, Foulds2010} is a variation on supervised learning, where each learning example contains a \textit{bag} of \textit{instances}. In MIL, a positive bag contains at least one positive instance. On the other hand, a negative bag contains no positive instances. Each audio clip in Audio Set contains several feature vectors. An audio clip is labelled positive for a class if at least one feature vector belongs to the corresponding class. 

A multi instance dataset consists of many pairs $ \{ B_{n}, d_{n} \}, n=1, ..., N $, where $ N $ is the number training pairs. Each bag $ B_{n} $ consists of several instances $ B_{n} = \{ x_{n1}, ..., x_{nL} \} $, where $ x_{nl} $ is an instance in a bag and $ L $ is the number of instances in each bag. We denote $ d_{n} $ as the label of the $ n $-th bag. In Audio Set classification, a bag is a collection of $ L $ features from an audio clip. Each instance $ x_{nl} \in \mathbb{R}^M $ is a feature, where $ M $ is the dimension of the feature. The label of a bag is $ d_{n} \in \{ 0, 1 \}^K $ where $ K $ is the number of audio classes and $ 0 $ and $ 1 $ represent the negative and positive label, respectively. For a specific class $ k $, when the label of the $ n $-th bag $ d_{nk}=1 $ then $ \exists x_{nl} \in B_{n} $ so that $ x_{nl} $ is positive. Otherwise if $ d_{nk}=0 $ then $ \forall x_{nl} \in B_{n} $ so that $ x_{nl} $ is negative. Assume we have a classifier $ f $ on each instance, we want to obtain a classifier $ F $ on each bag. There are several ways to obtain bag level classifier from instance level classifier described as follows. 

\subsection{Collective assumption}
The \textit{collective assumption} \cite{Xu2003} states that all instances in a bag contribute equally and independently to the bag's label. Under this assumption, the bag level classifier $ F $ is obtained by using the sum as the aggregation rule: 

\begin{equation} \label{eq1}
F(B) = \frac{1}{L} \sum_{x_{l} \in B} f(x_{l}). 
\end{equation}

\noindent The collective assumption is simple and assumes that the instances contribute equally and independently to the bag-level class labels. However the collective assumption assumes all the instances inherit the label from its corresponding bag, which is not the same as the MIL assumption. 

\subsection{Maximum selection}
The \textit{maximum selection} \cite{amores2013} states that the prediction of a bag is the maximum classification value of each instance in the bag described as follows: 

\begin{equation} \label{eq1}
F(B) = \underset{x_{l} \in B}{\max}f(x_{l}). 
\end{equation}

\noindent Maximum selection has been used in audio tagging using convolutional neural networks (CNNs) \cite{choi2016automatic} and audio event detection using weakly labelled data \cite{kumar2016audio}. Maximum selection corresponds to a global max pooling layer \cite{choi2016automatic} in a convolutional neural network. Maximum selection performs well in audio tagging \cite{choi2016automatic} but is sometimes inefficient in training because only one instance with the maximum value in a bag is used for training, and the gradient will only be computed from the instance with the highest classification value. 

\subsection{Weighted collective assumption}
The \textit{weighted collective assumption} is a generalization of the collective assumption, where a different weight $ w(x) $ is allowed for each instance $ x $ \cite{Foulds2010}:

\begin{equation} \label{eq1}
F(B) = \frac{1}{\sum_{x \in B} w(x)}\sum_{x \in B}w(x)f(x). 
\end{equation}

\noindent The weighted collective assumption asserts that each instance contributes independently but not necessarily equally to the label of a tag. This is achieved by incorporating a weight function $ w(x) $ into the collective assumption. Equation (3) has the same form as our joint detection-classification (JDC) model \cite{kong2017} and our attention model \cite{yongIS2017} proposed for audio tagging and sound event detection. The difference is that the work in \cite{kong2017, yongIS2017} model both $ w(x) $ and $ f(x) $ using neural network.

\section{Attention a probabilistic perspective} \label{sec:attention}
Although Equation (3) has been used in many previous works \cite{Foulds2010, kong2017, yongIS2017}, the explanation for this equation is not clearly presented. In this paper we explain this attention model in Equation (3) from a probabilistic perspective, which is helpful to guide the selection of $ f $ and $ w $ in Equation (3). 

\subsection{Measure space}
For any instances $ x $ in a bag, they should contribute differently to the classification of a bag. In MIL, a bag is labelled positive if at least one instance in the bag is positive. To solve this problem, the positive instances should be attended to and the negative instances should be ignored. We first assign a \textit{measure} on each $ x \in \Omega $ where $ \Omega $ is a set $ x $ laid in, for example Euclidean space. To assign the measure on each instance $ x $, we introduce the \textit{measure space} \cite{Chung2001} in probability theory. 

\textbf{Definition 1.} Let $ \Omega $ be a set, $ \mathscr{F} $ a Borel field \cite{Chung2001} of subsets of $ \Omega $. A \textit{measure} $ \mu $ on $ \mathscr{F} $ is a numerically valued set function with domain $ \mathscr{F} $, satisfying the following axioms:

\noindent 1. $ \forall E \in \mathscr{F}: \mu(E) \geq 0 $ \\
2. If $ \{E_{j}\} $ is a countable collection of disjoint sets in $ \mathscr{F} $, $ \mu(\bigcup_{j}E_{j}) = \sum_{j}\mu(E_{j}) $, then 
we call the triple $ (\Omega, \mathscr{F}, \mu) $ a \textit{measure space}. 

In addition, if we have:  \\
\noindent 3. $ \mu(\Omega) = 1 $ \\
\noindent then we call the triple $ (\Omega, \mathscr{F}, \mu) $ a \textit{probability space}. 

\subsection{Probability space}
When classifying a bag, different instances contribute differently. We define a probability space for each bag $ B_{n} $ for each class $ k $. As $ B_{n} \subset \Omega $, we may define a probability space $ (B_{n}, \mathscr{F}_{B_{n}}, p_{nk}) $ on $ B_{n} $ where $ \mathscr{F}_{B_{n}} = \mathscr{F} \bigcap \mathscr{F}(B_{n}) $ and $ \mathscr{F}(B_{n}) $ is the Borel filed of the set $ B_{n} $. The probability measure $ p_{nk} $ on $ B_{n} $ satisfies $ \sum_{x \in B_{n}} p_{nk}(x) = 1 $, so Definition 1 Axiom 3 is satisfied. We call $ (B_{n}, \mathscr{F} \bigcap B_{n} , p_{nk}) $ a probability space for the $ k $-th class. For an instance $ x $ in a bag, the closer $ p_{nk}(x) $ to 1 the more this instance is attended. The closer $ p_{nk}(x) $ to 0 the less this instance is attended. 

\subsection{Expectation}
Assume for the $ k $-th class, the classification prediction and the probability measure on each instance $ x \in B_{n} $ are $ f_{k}(x) $ and $ p_{nk}(x) $, respectively. To obtain the classification result on the bag $ B_{n} $, we apply the expectation of the classification result $ f_{k}(\cdot) $ with respect to the probability measure $ p_{nk} $:

\begin{figure}[t]
  \centering
  \centerline{\includegraphics[width=\columnwidth]{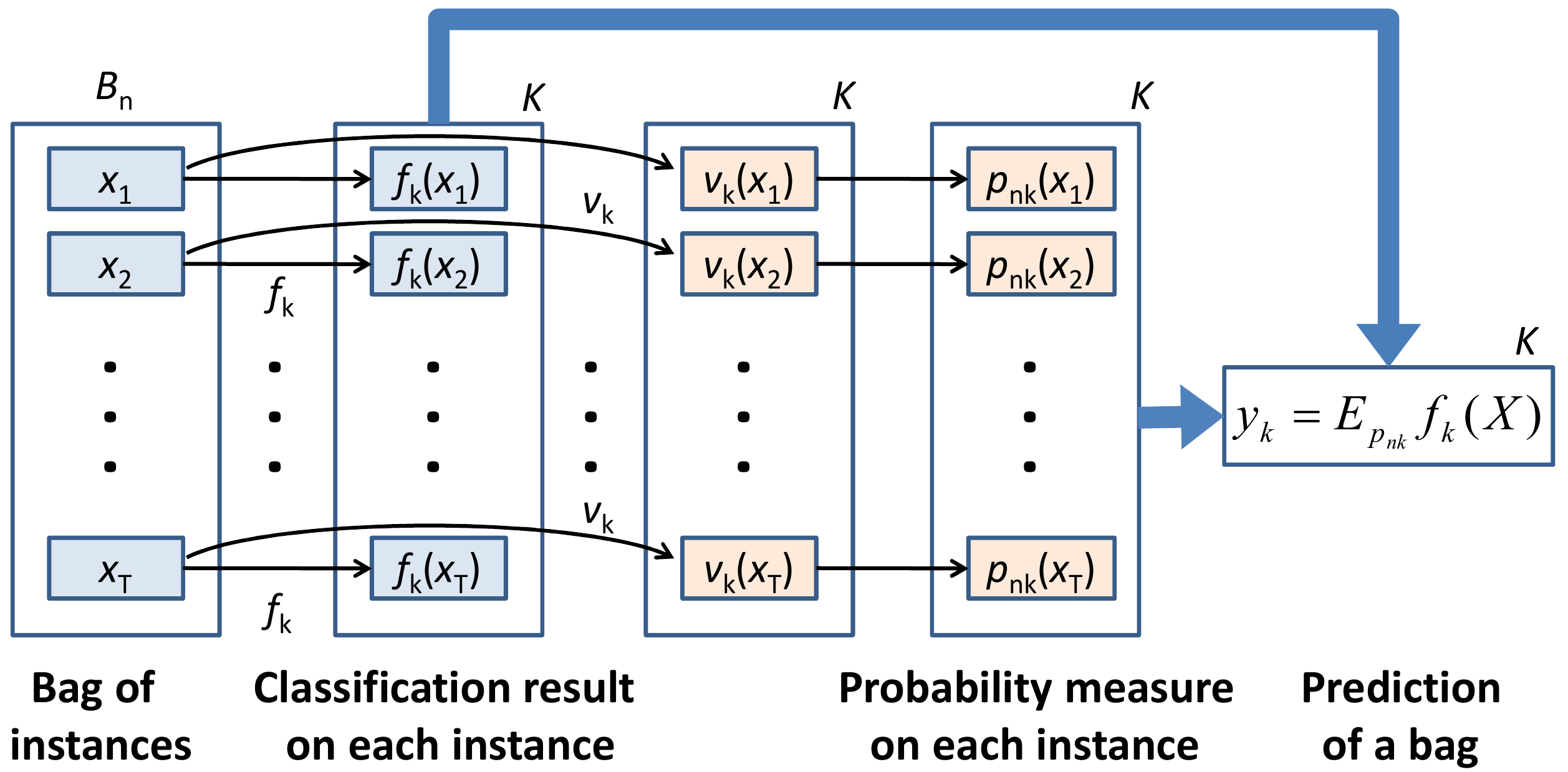}}
  \caption{Attention model a probabilistic perspective where $ f_{k} $ is the classification result on each instance and $ p_{nk} $ is the probability measure of each instance in a given bag. The prediction is the expectation of $ f_{k} $ with respect to the probability measure $ p_{nk} $. }
  \label{fig:results}
\end{figure}

\begin{equation} \label{eq1}
F(B_{n}) = E_{p_{nk}}(f_{k}(X)) = \sum_{x \in B_{n}} f_{k}(x)p_{nk}(x)
\end{equation}

\noindent where $ X $ is a random variable. Equation (4) represents the instances $ x \in B_{n} $ contributes differently to the classification of the bag $ B_{n} $. The probability measure $ p_{nk}(\cdot) $ controls how much an instance $ x $ is attended. Large $ p_{nk} $ and small $ p_{nk} $ represents the instance is attended and ignored, respectively.

\subsection{Modeling attention}
For a dataset with $ \Omega = \mathbb{R}^M $. A mapping $ f_{k}: \mathbb{R}^M \mapsto [0,1] $ is used to model the presence probability of the $ k $-th class of an instance $ x $. On the other hand, modeling the probability measure $ p_{nk}: \mathbb{R}^M \mapsto [0,1] $ is difficult because of the constraint that the sum of the probability of the instances in a bag should be equal to 1:

\begin{equation} \label{eq1}
\sum_{x \in B_{n}} p_{nk}(x) = 1. 
\end{equation}

\noindent So instead of modeling $ p_{nk} $ directly, we start from modeling $ \mu_{k} $ in the measure space $ (\mathbb{R}^M, \mathscr{F}, \mu_{k}) $ because in the measure space $ \mu_{k} $ does not need to satisfy Definition 1, Axiom 3. To model $ \mu_{k} $, we use a mapping $ v_{k}: \mathbb{R}^M \mapsto \overline{\mathbb{R}}^{+} $, where $ \overline{\mathbb{R}}^{+} = \mathbb{R}^{+} \bigcup \{0\} $. Then for each bag $ B_{n} $ and $ x \in B_{n} $, we may define the probability measure of any instance $ x $ of the $ k $-th class as:

\begin{equation} \label{eq1}
p_{nk}(x) = \mu_{k}(\{x\}) / \mu_{k}(B_{n}) = v_{k}(x) / \sum_{x \in B_{n}}v_{k}(x)
\end{equation}

\noindent where $ \mu(\{x\}) $ and $ \mu(B_{n}) $ are the measure of $ \{x\} $ and $ B_{n} $, respectively. From Definition 1 Axiom 2, $ \mu_{k}(B_{n}) $ can be calculated by $ \mu_{k}(B_{n}) = \sum_{x \in B_{n}} \mu_{k}(\{x\}) $. So the constraint in Equation (5) is satisfied. After modeling $ f_{k} $ and $ p_{nk} $, the prediction of the $ k $-th class can be obtained by using Equation (4). The framework of the attention model is shown in Fig. 1. 

\subsection{Mini batch balancing}
The Audio Set dataset is highly unbalanced. Some classes have tens of thousands samples while other classes only contain hundreds of samples. We therefore propose a mini batch balancing strategy, where the occurrence frequency of training samples of the different classes in a mini-batch are kept the same.

\section{Experiments}\label{sec:exp}
\subsection{Dataset}
We experiment on the Audio Set dataset \cite{gemmeke2017audio}. Audio Set contains over 2 million 10 seconds audio clips extracted from YouTube videos. Audio Set consists of 527 classes of audio with a hierarchy structure. The original waveform of the 2 million audio clips are not published. Instead, we use the published bottleneck feature vectors extracted from the embedding layer representation of a deep CNN trained on the YouTube-100M dataset \cite{youtube100m}. The bottleneck feature vectors are extracted at one feature per second, that is, there are 10 features in an 10 seconds audio clip. Then the bottleneck feature vectors are post-processed by a principle component analysis (PCA) to remove the correlations and only the first 128 PCA coefficients are kept. 

\subsection{Model}
The source code of this system is available here\footnote{\texttt{https://github.com/qiuqiangkong/ICASSP2018\textunderscore audioset}}. We apply a simple fully connected deep neural network to verify the effectiveness of the proposed attention model. We first apply fully connected layers on the input feature vectors to extract high level representation. We call this mapping as \textit{embedded mapping} and denote as $ g $. We call $ h=g(x) $ as \textit{embedded instance}. The embedded mapping $ g $ is modeled by three fully connected layers, with 500 hidden units in each layer followed by ReLU \cite{relu2010} non-linearity and dropout \cite{dropout2014} rate of 0.2 to reduce the risk of over-fitting. These configurations are chosen empirically. Then we model the classifier $ f_{k} $ and the measure $ v_{k} $ on each embedded instance $ h $ by the following equation:

\begin{equation} \label{eq1}
f_{k}(h) = \sigma(W_{f}h + b_{f})_{k}
\end{equation}
\begin{equation} \label{eq1}
v_{k}(h) = \phi(W_{v}h + b_{v})_{k}
\end{equation}

\noindent where $ \sigma $ is sigmoid non-linearity $ f(z) = 1 / (1 + e^{-z}) $. The sigmoid non-linearity ensures that the probability $ f_{k}(\cdot) $ is between 0 and 1. The non-linearity $ \phi $ can be any non-negative function and we investigate ReLU \cite{relu2010}, sigmoid and softmax functions in our experiment. 

Then we may obtain $ p_{nk} $ in the $ n $-th bag by:

\begin{equation} \label{eq1}
p_{nk}(x) = v_{k}(g(x)) / \sum_{x \in B_{n}}v_{k}(g(x))
\end{equation}

\begin{figure}[t]
  \centering
  \centerline{\includegraphics[width=\columnwidth]{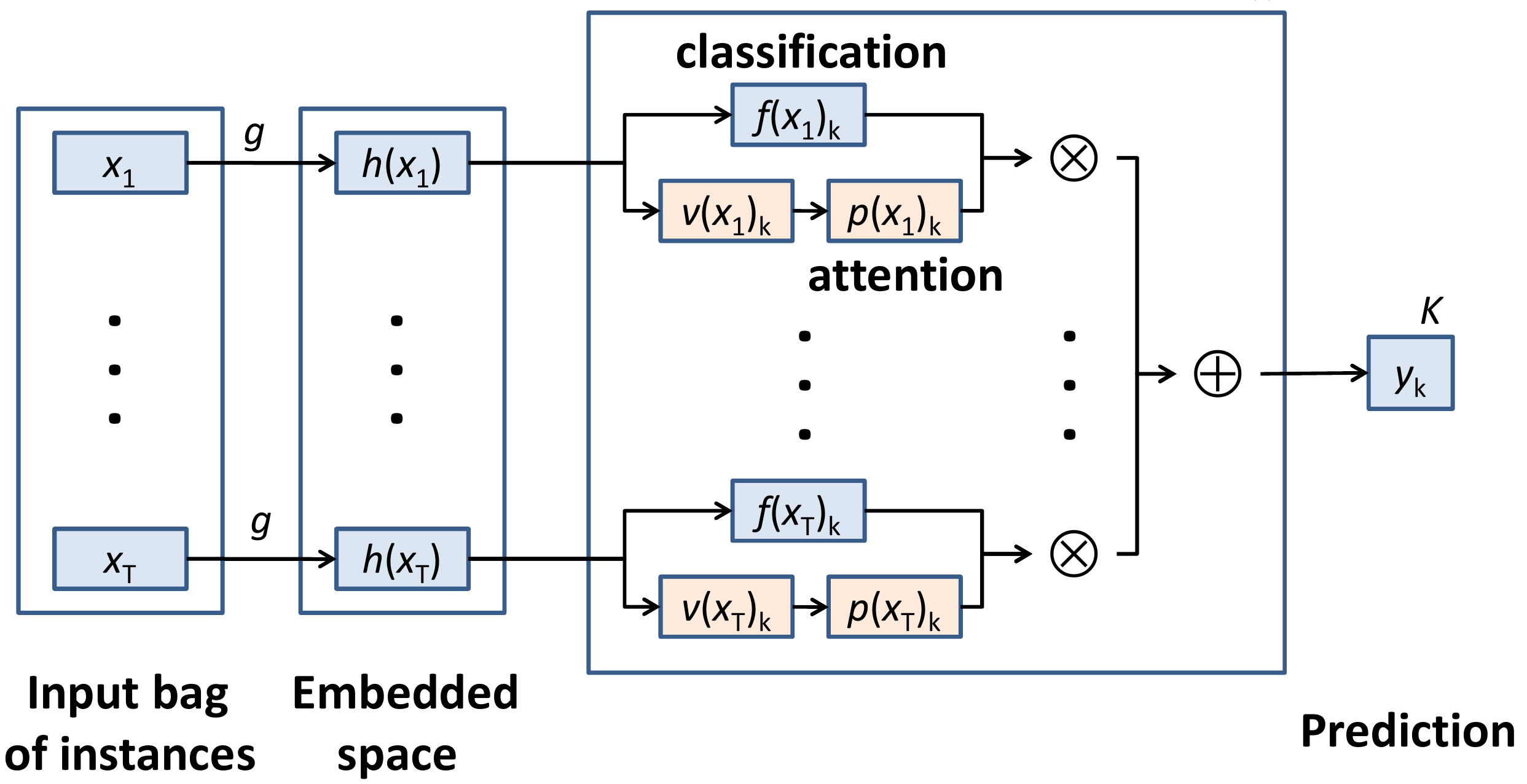}}
  \caption{Model for Audio Set classification. The input space is mapped to an embedded space followed by a classification and an attention (probability measure) branch. Final prediction is the expectation of the classification output with respect to the learned probability measure. }
  \label{fig:results}
\end{figure}

\noindent Finally the prediction of the $ k $-th event in bag $ B_{n} $ is obtained by using Equation (4). 

\subsection{Experiment analysis}
We evaluate using mean average precision (mAP), area under curve (AUC) and d-prime used in \cite{gemmeke2017audio}. These values are computed for each of the 527 classes and averaged across the 527 classes to obtain the final mAP, AUC and d-prime. Higher mAP, AUC and d-prime lead to better performance. 

Table 1 shows the results of with and without data balancing strategy using collective assumption in Equation (1). The data balancing strategy is described in Section 3.5. Table 1 shows using balancing strategy performs better than without data balancing strategy in all of mAP, AUC and d-prime. 

Table 2 shows the results of modeling the measure function $ v_{k}(\cdot) $ using different non-negative functions including ReLU, sigmoid and softmax functions. Softmax non-negative performs slightly better than sigmoid non-negative and better than ReLU non-negative function.  

Table 3 shows the comparison of different pooling strategies. Average pooling and max pooling along time axis are described in Equation (1) and (2), respectively. The Google baseline uses a simple fully connected DNN \cite{gemmeke2017audio}. Table 3 shows that RNN with global average pooling performs better than Google baseline. Using DNN with attention achieves better performance than Google baseline and RNN. 

\begin{table}[h]
\centering
\caption{Classification result with and without data balancing strategy. }
\resizebox{\columnwidth}{!}{%
\begin{tabular}{ p{3cm} p{2cm} p{2cm} p{2cm} }
 \hline 
 & mAP & AUC & d-prime \\
 \hline
 w/o balancing & 0.275 & 0.957 & 2.429 \\
 \hline
 with balancing & \textbf{0.296} & \textbf{0.960} & \textbf{2.473} \\
 \hline
\end{tabular}}
\end{table}

\begin{table}[h]
\centering
\caption{Classification results of measure $ v_{k}(\cdot) $ modeled by ReLU, sigmoid and softmax functions.  }
\resizebox{\columnwidth}{!}{%
\begin{tabular}{ p{3.5cm} p{2cm} p{2cm} p{2cm} }
 \hline 
 & mAP & AUC & d-prime \\
 \hline
 DNN ReLU attention & 0.306 & 0.961 & 2.500 \\
 DNN sigmoid attention & 0.326 & 0.964 & 2.547 \\
 DNN softmax attention & \textbf{0.327} & \textbf{0.965} & \textbf{2.558} \\
 \hline
\end{tabular}}
\end{table}

\begin{table}[h]
\centering
\caption{Classification results with different pooling strategy. }
\resizebox{\columnwidth}{!}{%
\begin{tabular}{ p{3.5cm} p{2cm} p{2cm} p{2cm} }
 \hline 
 & mAP & AUC & d-prime \\
 \hline
 DNN max pooling & 0.284 & 0.958 & 2.442 \\
 DNN avg. pooling & 0.296 & 0.960 & 2.473 \\
 Google baseline & 0.314 & 0.959 & 2.452 \\
 RNN avg. pooling & 0.325 & 0.960 & 2.480 \\
 DNN softmax attention & \textbf{0.327} & \textbf{0.965} & \textbf{2.558} \\
 \hline
\end{tabular}}
\end{table}

\section{Conclusion}\label{sec:conc}
In this paper, an attention model in audio classification is explained from a probability perspective. Both the classifier and the probability measure on each instance are modeled by a neural network. We apply fully connected neural network with this attention model on Audio Set and achieves mAP of 0.327 and AUC of 0.965 outperforming the Google baseline and recurrent neural network. In the future, we will explore more on modeling probability measure using different non-negative functions. 

\section{Acknowledgement}
This research is supported by EPSRC grant EP/N014111/1 ``Making Sense of Sounds'' and Research Scholarship from the China Scholarship Council (CSC). 

\vfill\pagebreak

\bibliographystyle{IEEEbib}
\bibliography{refs}

\end{document}